\documentclass[runningheads,fleqn]{svmult}

\usepackage{makeidx}         
\usepackage{multicol}        
\usepackage{subeqnar}  
\usepackage{taphys}    
\usepackage{graphicx}        
\makeindex             

\newcommand{\beq}{\begin{equation}}
\newcommand{\eeq}{\end{equation}}
\newcommand{\bea}{\begin{eqnarray}}
\newcommand{\eea}{\end{eqnarray}}
\newcommand{\pdag}{{\phantom{\dagger}}}
\newcommand{\bk}{{\bf k}}

\begin{document}

\title*{Flavor Degeneracy and Effects of Disorder in Ultracold Atom Systems}
\toctitle{Flavor degeneracy and effects of disorder in ultracold atom systems}
\titlerunning{Flavor degeneracy and effects of disorder in ultracold atom systems}
\author{Walter Hofstetter}
\authorrunning{Walter Hofstetter}
\institute{Institut fuer Theoretische Physik A, RWTH Aachen,  Templergraben 55,
\mbox{52056 Aachen}, Germany}

\maketitle

\begin{abstract}
Cold atoms in optical lattices offer an exciting 
new laboratory where quantum many-body phenomena 
can be realized in a highly controlled way.  
They can serve as \emph{quantum simulators} for 
notoriously difficult problems like high-temperature 
superconductivity. 
This review is focussed on recent developments  
and new results in multi-component systems.   
Fermionic atoms with SU(N) symmetry have exotic superfluid 
and flavor-ordered ground states. We discuss symmetry breaking, 
collective modes and detection issues. 
Bosonic multi-flavor ensembles allow for  
engineering of spin Hamiltonians which are interesting 
from a quantum computation point of view. 
Finally, we will address the competition of disorder and interaction in optical lattices.  
We present a complete phase diagram obtained within 
dynamical mean-field theory and discuss experimental 
observability of the Mott and Anderson phases. 
\end{abstract}

\section{Introduction and Overview \label{sec:1}}
%







%
The achievement of Bose-Einstein condensation (BEC) 10 years ago \cite{bec} 
has opened the new field of interacting quantum gases in the dilute limit. 
It has become possible to observe quantum phenomena like Bose statistics 
on a mesoscopic scale, involving a large number of atoms. 
More recently, fermionic gases have also been cooled to the quantum 
degenerate regime, using sympathetic cooling of two spin states or boson-fermion mixtures 
\cite{deMarco,truscott,fer3,fer4}. Although the resulting temperatures $T/T_F \approx 0.1$ 
are, relatively to the Fermi temperature $T_F$, much higher than in solids, the Pauli principle  
has been clearly observed. 

In addition to quantum statistics, tunable interactions are another important ingredient 
in the cold atom "toolbox". The interactions between atoms can be changed by an external magnetic field as a result of 
Feshbach resonances \cite{inouye,timm}. In particular, their scattering length can be tuned to 
positive or negative values, corresponding to repulsive or attractive interactions. 
This has opened the way to studies of solid-state related phenomena like Cooper pairing 
and BCS superfluidity of fermions \cite{holland,ohashi}. 
The resulting BEC-BCS crossover has recently been the subject of intense experimental 
and theoretical studies \cite{regal04,zwierlein04,grimm04} 

In an independent development, degenerate atomic clouds have been combined 
with optical lattices, created by standing light waves which generate an effective periodic 
potential for the atoms \cite{orzel,greiner,boseth}.  Interactions  can thus be tuned without changing 
the atomic scattering length. This has been demonstrated in a pathbreaking experiment 
\cite{greiner} where interacting bosons were tuned through a quantum phase transition 
from a superfluid (SF) to a Mott insulating state. 
Very recently, fermionic K$^{40}$ atoms have been loaded into 3d optical lattices as well \cite{esslinger}. 
In these new experiments the lowest Bloch band was filled up succesively, and the shape of the Fermi surface 
monitored by time-of-flight measurements. Eventually a completely filled Brillouin 
zone corresponding to a band insulator was observed. 

More generally, fermionic atoms in optical lattices allow for the realization of solid-state type quantum phases 
like antiferromagnetism or high-temperature superconductivity \cite{high_temp_sf}.  
Even the spatial dimensionality of the lattice can be tuned. As an example, one-dimensional optical lattices have been 
realized where the hardcore or Tonks-gas limit of interacting bosons has been observed \cite{cazalilla,tonks,stoeferle}. 
Recent progress in numerical methods for simulating 1d quantum systems has lead 
to interesting predictions about the dynamics of such systems \cite{Kollath04a,Kollath04b}. 

In the following we will first introduce the basic model describing cold atoms in optical lattices.  
We will then address systems with multiple flavors, i.e. hyperfine states, which 
allow realization of new exotic quantum states not accessible in solids. 
Finally, we will discuss the role of disorder in current and future experiments involving cold atoms.

\section{Optical Lattices and Strong Correlations}
\label{sec:2}







\subsection{Model and Parameters}
\label{sec:parameters}

\begin{figure}
\begin{center}
\includegraphics[width=0.5\linewidth]{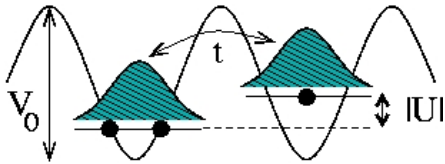}
\end{center}
\caption{\label{fig:1} Cold atoms in an optical lattice of strength $V_0$, shown here 
with hopping $t$ and negative onsite interaction $U$. This situation corresponds to an attractive 
Hubbard model where multiple occupancy of lattice sites is energetically favourable.}
\end{figure}

Atoms can be trapped in standing light waves 
created by interfering laser beams detuned far from resonance \cite{orzel,greiner,boseth}.
Due to the AC Stark shift the atoms experience a periodic potential of the form 
$ V(x) = V_0 \sum_{i=1,2,3} \cos^2 (k x_i) $
where $k$ is the wave vector of the laser. The natural energy scale for the potential depth $V_0$ is 
the recoil energy  $E_R = \hbar^2 k^2 / 2 m$. 
A schematic picture of such an optical lattice is shown in Fig.~\ref{fig:1}. 
The eigenstates in the periodic lattice potential  are given by Bloch bands, and 
an equivalent representation  in terms of Wannier orbitals   
leads to a tight-binding Hamiltonian. 
Let us assume for the moment that two different (hyperfine-) spin states are present, which 
in the following are denoted as $\sigma = \uparrow, \downarrow$.  
If temperature and filling are sufficiently low, the atoms will be confined exclusively to the lowest Bloch band. 
In this case the system can be described by a Hubbard Hamiltonian \cite{hubbard,boseth}
\beq
H = -t \sum_{<i j>, \sigma} 
\left(c^\dagger_{i \sigma} c^\pdag_{j \sigma} + c^\dagger_{j \sigma} c^\pdag_{i \sigma}\right) 
+ U \sum_i n_{i \uparrow} n_{i \downarrow}
\label{Hubbard}
\eeq
where $\langle i j \rangle$ labels next neighbors, 
$c_{i \sigma}$ is the fermionic annihilation operator for the Wannier state of spin $\sigma$ on site $i$ 
and $n_{i \sigma} = c^\dagger_{i\sigma} c^\pdag_{i\sigma}$ is the corresponding number density. 
The parameters for hopping  and onsite interaction  can then be expressed as 
$t = E_R (2 / \sqrt{\pi}) \xi^3 \exp(-2 \xi^2)$ and 
$U = E_R a_s k \sqrt{8/\pi} \, \xi^3 $. 
Here $a_s$ is the atomic scattering length and 
$\xi = (V_0 / E_R)^{1/4}$ is a parameter characterizing the strength of the lattice 
\cite{boseth,high_temp_sf}.
It is obvious that by tuning the optical lattice potential $V_0$ 
one can achieve arbitrary ratios $|U|/t$ without changing $a_s$. 
Optical lattices thus give access to the \emph{strongly correlated} regime without using 
Feshbach resonances. 

\begin{figure}
\begin{minipage}{0.48\linewidth}
\begin{center}
\includegraphics[width=0.7\linewidth]{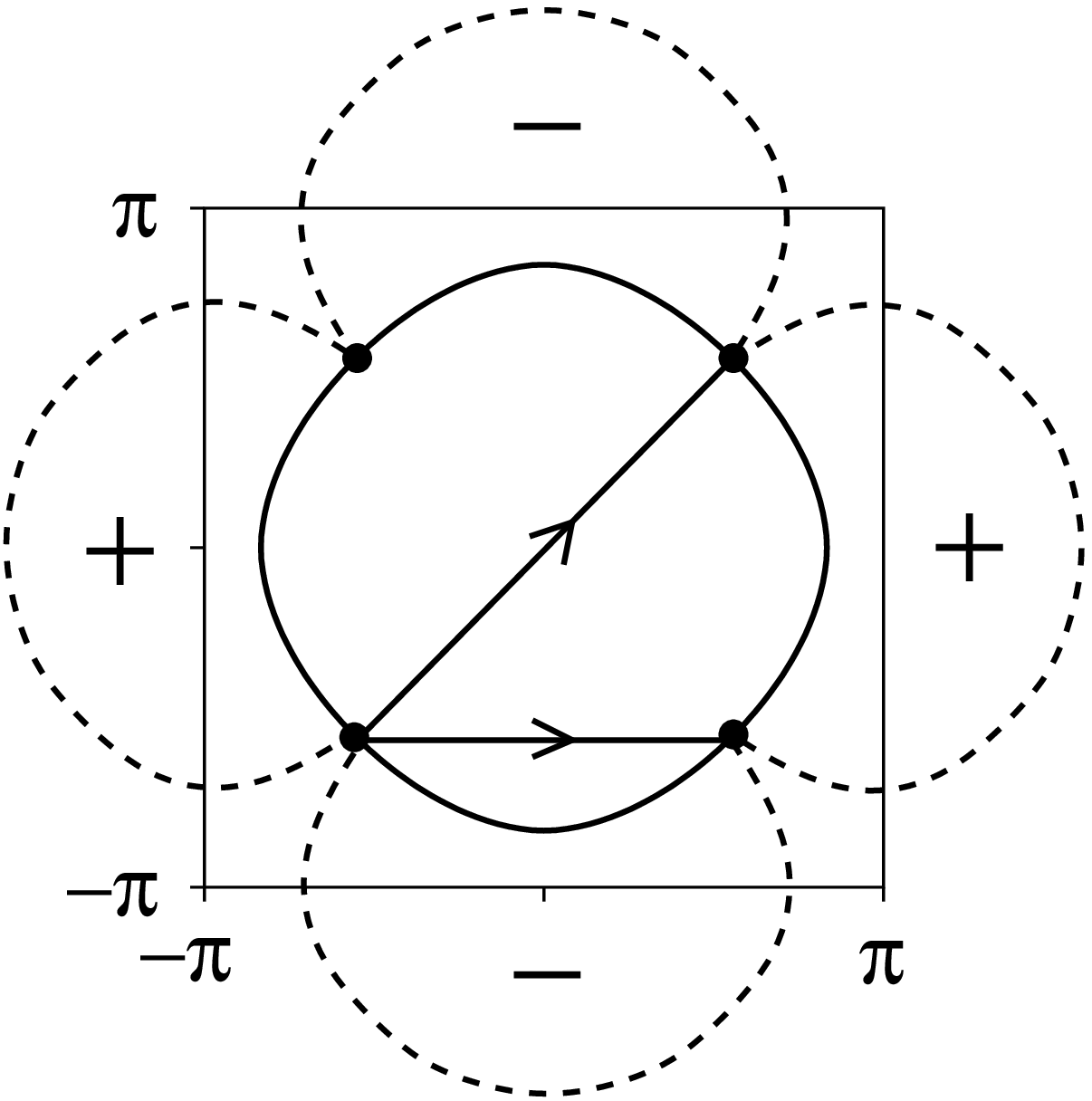}
\end{center}
\end{minipage}
\hfill
\begin{minipage}{0.48\linewidth}
\begin{center}
\includegraphics[width=0.7\linewidth]{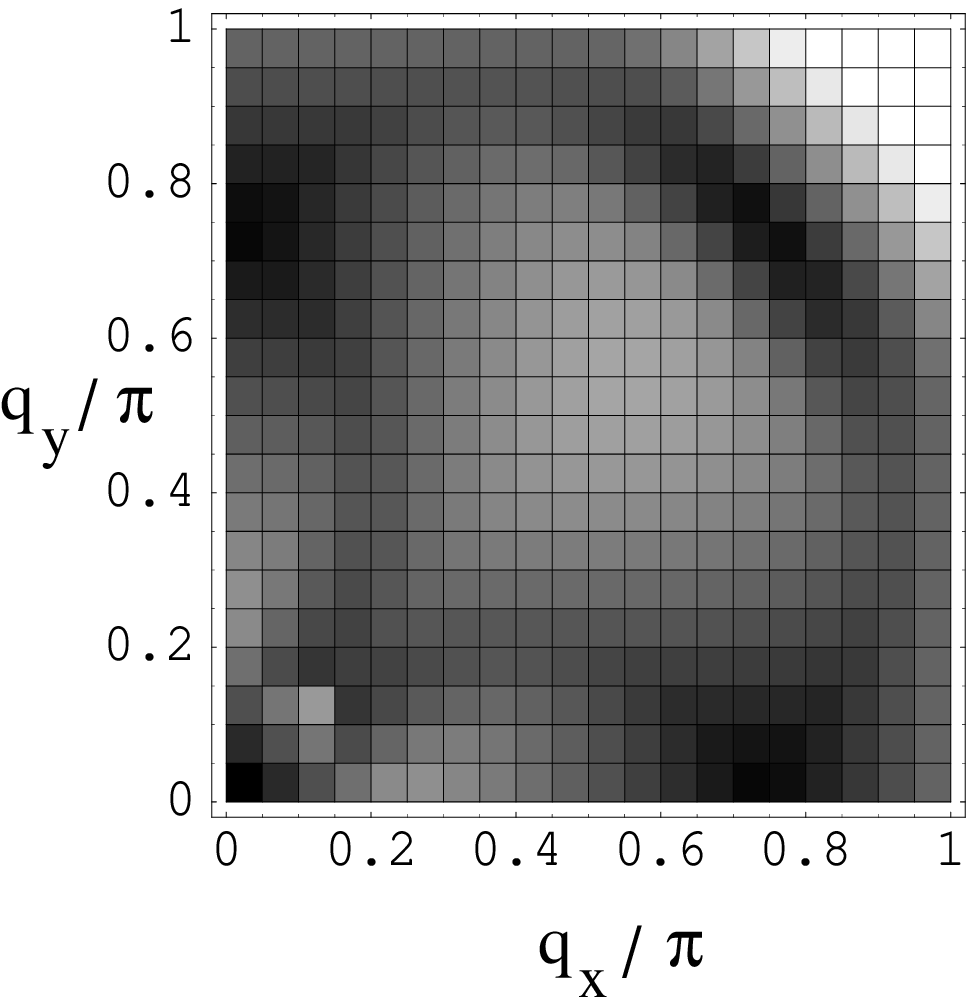}
\end{center}
\end{minipage}
\caption{\label{fig:3} Probing $d$-wave pairing in the repulsive 2d Hubbard model via Bragg scattering. 
Left: schematic diagram of the Fermi surface in 2d (solid line) and the momentum dependence of the gap 
(dashed line). 
Right: onset frequency of the quasiparticle continuum in the dynamical structure factor $S(q, \omega)$, 
plotted as a function of momentum $q$. At the special wave vectors connecting the nodal points 
in the left figure, the density response is gapless.
Figures taken from \cite{high_temp_sf}. }
\end{figure}

\subsection{Quantum Simulations \label{sec:high_TC}}
The Hubbard model (\ref{Hubbard}) is of fundamental importance for electronic correlation effects in condensed matter. 
From this point of view, ultracold atoms can be used to perform \emph{quantum simulations} 
of solid state physics. Here we illustrate this intriguing idea with the example of high-temperature 
d-wave superconductivity \cite{high_temp_sf}.

Consider the 2d Hubbard Hamiltonian (\ref{Hubbard})  with spin-1/2 fermions and repulsive interaction $U>0$ 
resulting from a positive scattering length $a_s > 0$. At half filling 
$n_{i} = 1$ this model gives rise to staggered antiferromagnetic order. At lower filling fractions, 
theoretical arguments suggest a d-wave paired phase, 
which is a possible candidate for explaining high-temperature superconductivity in the cuprates 
\cite{dwave}. However, there is no satisfactory numerical evidence, mainly because quantum Monte Carlo 
calculations are limited to extremely small systems due to the sign problem. 

On the other hand, cold fermions in an optical lattice could be used to \emph{experimentally} probe  
d-wave pairing in the 2d Hubbard model. 
The resulting superfluid order can be detected via Bragg 
scattering, as shown in Fig.~\ref{fig:3}, which is by now a well-established technique 
to measure the dynamical density response $S(q, \omega)$ in interacting quantum gases \cite{ketterle}.
As already suggested by Feynman \cite{feynman}, 
such quantum simulations  could 
provide a powerful tool to gain insight into 
many-body Hamiltonians relevant for solid-state physics.







\section{Multi-Component Systems}
\label{sec:3}






\subsection{Two-Component Bosons with Spin Order}

All of the alkali atoms available for trapping and cooling have $2*(2 I + 1)$ low-lying 
hyperfine states, where $I$ is the nuclear spin. 
Several of these states can be trapped at the same time: in magnetic traps one is limited 
by the condition that the states have to be \emph{low-field seekers},  
but optical dipole traps, as well as optical lattices, 
allow confinement of basically any combination of spin states \cite{optical_trap}, 
as long as no instability due to three-body collision occurs.  
Loading a lattice with two hyperfine states of Rb$^{87}$ has been demonstrated 
experimentally in \cite{mandel} 
where also a spin-dependent periodic potential has been implemented. 
In the following we discuss a proposal, described in detail in \cite{2comp_mott}, 
how these techniques can be used to engineer quantum spin Hamiltonians which in turn 
could be relevant for quantum information processing. 

Let us consider a system of two bosonic hyperfine states in a lattice, described by the following 
Bose-Hubbard Hamiltonian:
\bea
H &=&- t_a \sum_{\langle ij\rangle}  \left(a^\dagger_i a^\pdag_j +H.c \right)
-t_b\sum_{\langle ij\rangle } \left(b^\dagger_i b^\pdag_j+H.c\right) 
+ U\sum_i (n_{ai} - {1 \over 2}) (n_{bi} - {1 \over 2}) \nonumber \\
& & +{1 \over 2} \sum_{i, \alpha =a,b}V_\alpha n_{\alpha i}(n_{\alpha i}-1)
-\sum_{i, \alpha}\mu_\alpha n_{\alpha i}. 
\label{BHM}
\eea
Here $a_i, b_i$ denote the annhilation operators for two different bosonic pseudospin states, 
and the number operators are defined as $n_{a i} = a^\dagger_i a^\pdag_i$, $n_{b i} = b^\dagger_i b^\pdag_i$,  
with the corresponding chemical potentials $\mu_{a (b)}$. 
In reality, experiments are performed at a fixed numbers of particles, which 
in the grand canonical description can be achieved by tuning the chemical potential. 
The onsite interaction beween equal spin states is given by $V_{a (b)}$, and the one between  
different spins by $U$. We also assume a spin dependent tunable hopping $t_{a (b)}$ 
which has already been experimentally realized \cite{mandel}. 

We now focus on the case of integer filling $n_a + n_b = 1$, following \cite{2comp_mott}.  
We are mainly interested in the nature of the Mott-superfluid transition in this system, 
and the possibility of spin order in the insulating phase. 
To address the second issue, it is instructive to first consider parameters \mbox{$t_{a, b} \ll U, V_{a, b}$} 
deep inside the Mott phase. States with double occupancy per site are then very unfavourable 
and can be projected out by a Schrieffer-Wolff transformation. This leads to an effective spin Hamiltonian 
in the subspace of single occupation \cite{duan}.  
The physics of this XXZ model is well understood and includes an x-y ferromagnetic phase 
as well as an antiferromagnetic z-Neel ordered state.  

\begin{figure}
\begin{minipage}{0.45\linewidth}
\includegraphics[width=\linewidth]{hofsfig4.eps}
\end{minipage}
\hfill
\begin{minipage}{0.45\linewidth}
\includegraphics[width=0.99\linewidth]{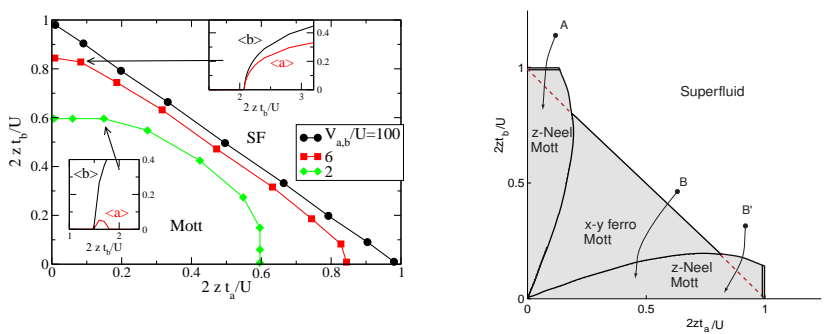} \\
\end{minipage}
\caption{\label{fig:4} Left: Phase diagram of the 2-component bosonic Hubbard model  
obtained by mean-field theory. Note that as $V_{a (b)}$ decreases, the Mott domain shrinks. 
Right: Phase diagram including quantum fluctuations.    
Figures taken from \cite{2comp_mott}. }
\end{figure}

The disadvantage of this deep Mott regime is that the critical temperature for magnetic ordering 
is very low $T_c \sim \max(t_{a (b)}^2 / U, \, t_{a(b)}^2/V_{a(b)})$ and therefore experimentally hardly accessible. 
In order to enhance $T_c$ and study the region close to the Mott-SF transition 
is necessary to make at least one of the interaction parameters 
comparable to the hopping. Here we choose $t_{a (b)} \approx U \ll V_{a (b)}$, which means 
that double occupancy with two different spins is now possible. 
The main question is whether the spin order discussed above is still visible close to the superfluid. 
In order to map out the Mott-SF phase boundary, we have used a mean-field approach 
first proposed in \cite{mott-mft}, where the kinetic energy is decoupled. 

The phase diagram obtained in this way is shown in Fig.~\ref{fig:4} (left). 
Note that as $V_{a (b)}$ decreases, the Mott domain shrinks.  
In order to resolve different spin states in the insulator, 
it is necessary to take into account quantum fluctuations 
on top of the variational mean-field state and compare the resulting ground state energies. 
Details of this calculation can be found in \cite{2comp_mott}. 
The resulting phase diagram including fluctuations is shown in Fig.~\ref{fig:4} (right). 
Spin ordering persists right up to the SF phase boundary and and can  
be tuned from xy-ferromagnetic to z-Neel antiferromagnetic by the ratio $t_a / t_b$. 
We find hysteresis between the Neel state and the superfluid, while the transition 
between the xy-state and the SF is continuous. 
These should be clear signatures for an experimental detection of spin ordered states, 
using for example Rb$^{87}$ atoms. 
The spin order can be directly observed by spin-dependent Bragg scattering 
or via density fluctuations in time-of-flight measurements \cite{ehud-noise}.

\subsection{Beyond Solid-State: SU(N) Fermions}

We will now show that with the degrees of freedom offered by ultracold atoms 
it is possible to create new states of matter that have no equivalent in condensed matter.  
The obvious constraint in solid-state physics is that electrons have only two spin states. 
Atoms, on the other hand, have large hyperfine multiplets out of which several states 
can be trapped simultaneously. 
For fermionic atoms this has been demonstrated with the three states 
$|F=9/2, m_F = -5/2, -7/2, -9/2 \rangle$ of K$^{40}$ in an optical trap \cite{regal}. 
Alternatively, one could use the three spin polarized $m_s = 1/2$ states of Li$^6$ 
which, in a sufficiently large field, have a pairwise equal and anomalously large 
triplet scattering length $a_s = -2160 a_0$ \cite{abraham}. 

\begin{figure}
\begin{center}
\includegraphics[width=0.7\linewidth]{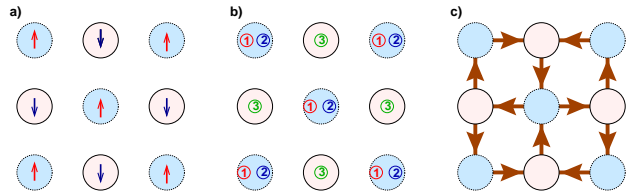}
\end{center}
\caption{\label{fig:5} 
Types of order in the $U>0$ fermionic SU(3) Hubbard model. 
a) AF spin-density wave for $N=2$.     
b) Flavor-density wave state for N=3. Flavor 1 and 2 prefer one sublattice, 
flavor 3 the other. 
c) Staggered flux state for $N>6$: particle currents are indicated by arrows.  
Figures taken from \cite{sun_prl}. }
\end{figure}

These systems can be used to realize fermionic Hubbard models with $N>2$ flavors and 
approximate SU(N) flavor symmetry. In the following we will discuss the rich physics of 
these models for finite $N$, following \cite{sun_prl,sun_prb}. The Hamiltonian is given by 
\begin{equation} 
H =  -t \sum_{m,\langle ij \rangle} 
    \left[c^\dagger_{i,m} c_{j,m} + c^\dagger_{j,m} c_{i,m} \right] + 
\frac{U}{2} \sum_i n_i^2   
\label{hubb}
\end{equation} 
where $c^\dagger_{i m}$ creates a fermion of flavor $m=1, \ldots N$ on site $i$ 
and $n_i = \sum_m n_{i, m}$ is the total number of atoms on site $i$. 
Note that the interaction term has local SU(N) invariance while the 
hopping reduces this to a global one. The values of $t$ and $U$ can be derived from 
atomic parameters along the lines of section \ref{sec:parameters}. 

While the large-$N$ limit of this model has been well studied in the context of 
high-$T_c$ superconductivity \cite{marston}, few results have been previously obtained for finite $N$. 
Consider first the case of repulsive interactions $U>0$. 
We have performed a systematic analysis of weak-coupling instabilities 
using a perturbative functional renormalization group (RG) approach \cite{tflow}.  
Although the RG eventually breaks down at strong coupling, it allows to identify the leading instability 
towards an ordered phase. The analysis performed in \cite{sun_prl} focusses on $d=2$ dimensions.

In Fig.~\ref{fig:5} the three relevant types of order at half filling $\langle n_i \rangle = N/2$ 
are shown. In the spin $1/2$ case the system displays staggered antiferromagnetic order, 
as is well known. For intermediate $N<6$ the RG yields an instability towards 
flavor density wave states with ordering wavevector ${\bf Q} = (\pi, \pi)$ like in the 
antiferromagnetic case. This corresponds to a breaking of the SU(N) symmetry, 
with a degenerate ground state manifold.  
As $N$ increases, breaking of SU(N) becomes less favorable because the 
number of Goldstone modes increases. For $N>6$ the RG indicates a dominant instability of the 
\emph{staggered flux} type with alternating particle currents around the plaquettes of the 
2d lattice (see Fig.~\ref{fig:5}c). 

Let us briefly discuss the temperature scales $T_c$ below which the respective long range orders set in. 
The critical temperature for flavor density waves at strong coupling scales like $t^2/U$  
and can thus be tuned to relatively large values: for $N=3$ the RG predicts a 
maximum of $T_c \approx 0.1 t$. 
On the other hand, staggered flux order, like d-wave superconductivity away from half filling, 
requires significantly lower temperatures, with a typical RG estimate given by $T_c \approx 0.01 t$ 
for $N=7$. This is about an order of magnitude below the current experimental limit 
and will require improved cooling techniques.  

\begin{figure}
\begin{center}
\includegraphics[width=0.7\linewidth]{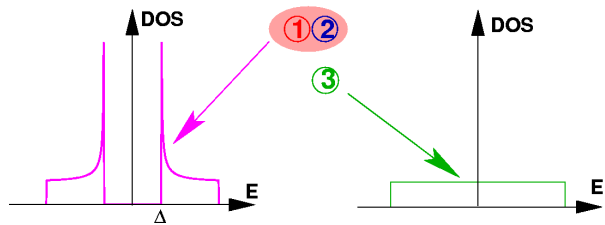}
\end{center}
\caption{\label{fig:6} 
BCS pairing of 3-flavor fermions with SU(3) symmetry. Note that one flavor remains unpaired, 
with a normal Fermi surface.  
Figure taken from \cite{sun_prb}. }
\end{figure}

Next, we focus on the situation with attractive interactions $U<0$ and $N=3$ flavors which is  relevant for Li$^6$. 
A large recent experimental effort has been devoted to the BEC-BCS crossover in spin-$1/2$  superfluid fermions \cite{regal04,zwierlein04,grimm04,formation_time}. 
A common feature of these experiments with K$^{40}$ and Li$^6$ is the use of 
a Feshbach resonance to generate large attractive interaction.  
These resonances generally occur only between two hyperfine spin states and thus cannot be used 
to realize an SU(3) symmetric model. 
However, as pointed out above,  Li$^6$ has a remarkably large and negative 
background scattering length which in a finite magnetic field is approximately 
equal for the three spin states with $m_s = 1/2$. In combination with an optical lattice one can thus 
realize the SU(N) Hubbard model (\ref{hubb}) with $U<0$ and $N=3$. 
The possibility of a three-flavor paired state in Li$^6$, without consideration of the SU(3) symmetry, 
has already been pointed out by Leggett \cite{modawi}. 

Following the analysis in \cite{sun_prb} we now discuss how the spin-$1/2$ BCS state 
is generalized to three flavors. 
We introduce a pairing mean-field and Hamiltonian 
\beq
\Delta_{\alpha \beta} = - {U \over N} \sum_{\bf k} \langle c_{\bk \alpha} c_{-\bk \beta} \rangle \ \ \ \ \ \ \ \ 
H_{\mathrm{MF}}= -\frac{1}{2} \sum_{\vec{k}, \alpha ,\beta}  
c^\dagger_{\vec{k}\alpha}   c^\dagger_{-\vec{k}\beta} \Delta_{\beta 
\alpha} + h.c.
\eeq
where $\alpha, \beta = 1, \ldots 3$ are flavor indices and $N$ is the number of lattice sites. 
We focus on s-wave pairing which is favorable for strong onsite attraction. 
The Pauli principle then requires antisymmetry $\Delta_{\alpha \beta} = - \Delta_{\beta \alpha}$ 
in the flavor index. 
The order parameter can therefore been written as a triplet 
$ \Delta_\alpha = {1 \over 2} \, \epsilon_{\alpha \beta \gamma} \langle c_\beta c_\gamma \rangle  = 
( \Delta_{23},  - \Delta_{13},  \Delta_{12} )$. 
From mean-field theory we obtain that all ground states consistent with 
$\sum_\alpha |\Delta_\alpha|^2 = \Delta_0^2 $ are degenerate. 
This five-dimensional ground-state manifold is consistent with the number of collective modes 
obtained via Goldstone's theorem. 
%

\begin{figure}
\begin{center}
\includegraphics[width=0.5\linewidth]{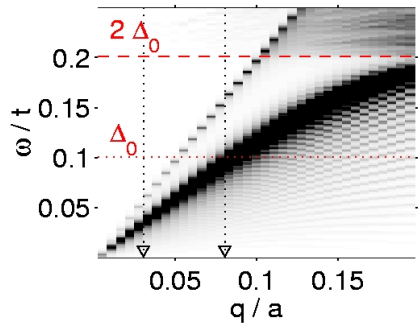}
\end{center}
\caption{\label{fig:7} 
Density response spectrum $\chi^{\rho}(q, \omega)$ 
(equivalent to $S(q, \omega)$) 
of the 2d fermionic SU(3) Hubbard model  
at $T = 0.01t$, $U = -4 t$ and filling $n \approx 0.55$.  
The Anderson-Bogoliubov mode, the signature of superfluidity, is clearly visible 
(thick black line) as well as an additional flavor mode indicating the 3-flavor degeneracy. 
Figure taken from \cite{sun_prb}. }
\end{figure}

The remarkable feature of this triplet s-wave state is that superfluid Cooper pairs coexist with 
a normal Fermi surface (see Fig.~\ref{fig:6}), i.e. the single-particle spectrum is only 
partially gapped. 
This has consequences for the collective modes which we have  
analyzed within a generalized RPA scheme \cite{sun_prb}. 
They are partially visible in the dynamical structure factor $S(q, \omega)$, which is accessible via 
Bragg scattering \cite{ketterle}. An example of the calculated density response  
${\rm Im} \chi^\rho(q, \omega)$ is shown in Fig.~\ref{fig:7}. 

From BCS mean-field theory in two dimensions we find a transition temperature of 
$T_c = 0.17 t$ for typical parameters $n = 3/8$ and $U = -4 t$. 
This amounts to roughly $0.05 T_F$ and is within reach of present cooling techniques. 
Multi-component Fermi systems like Li$^6$ can thus provide exotic new many-body physics 
and may even allow quantum simulations of simplified QCD models where only the 
color degree of freedom is taken into account.

\section{Disorder and Interaction}
\label{sec:4}

So far in this review we have focussed on pure, translationally invariant quantum lattice models. 
It is indeed one of the main advantages of optical lattices that perfectly disorder--free systems 
can be realized. On the other hand, effects of impurities and defects are of central importance in solids, 
where they often compete with the electron-electron interaction \cite{Lee85,Belitz94}. 
It is therefore of great interest to realize in a controlled 
way disordered cold atom systems where localization effects can be studied. 

Experimentally, disordered potentials can be created either with speckle lasers 
\cite{Horak98} or via quasiperiodic optical lattices \cite{Guidoni97}.  
Due to the AC stark effect the atoms experience a spatially fluctuating random potential which is 
stationary in time. Recently, localization effects have been observed in a BEC subject 
to a speckle laser field \cite{Lye04}. 

Here we focus on \emph{fermionic} atoms with two spin states in a 
three-dimensional optical lattice with an additional 
random potential. A complete presentation of the results discussed here can be found in \cite{anderson-mott}.  
The system is modelled by the Anderson-Hubbard Hamiltonian 
\begin{equation}
H_{\rm AH}=-t\sum_{\langle ij\rangle \sigma }c_{i\sigma }^{\dagger }c_{j\sigma
}+\sum_{i\sigma }\epsilon _{i}n_{i\sigma }+U\sum_{i}n_{i\uparrow
}n_{i\downarrow } - \mu \sum_{i \sigma} n_{i \sigma},  
\label{AH}
\end{equation}
where $\epsilon_i$ is a random onsite potential which we assume to be uniformly distributed 
in the interval $[-\Delta/2, \Delta/2]$. The parameter $\Delta$ is a measure of the disorder strength. 
We focus on the case of half filling $n = 1$ where on average there is one particle per site. 
The Hamiltonian (\ref{AH}) describes both the interaction-induced Mott transition into a correlated 
insulator \cite{Mott49,mott-quarter-fill} as well as the Anderson localization transition due to coherent backscattering 
from random impurities \cite{Anderson58}. 

Analyzing the model (\ref{AH}) is a challenging problem. 
Here we present results obtained within the \emph{Dynamical Mean-Field Theory} (DMFT), 
a nonperturbative technique where local quantum fluctuations are treated exactly \cite{Metzner89,Georges96}. 
The DMFT has been applied with great success in $d=3$ spatial dimensions 
to explain magnetic ordering phenomena and the Mott transition. In the calculation presented here \cite{anderson-mott} we use a recently developed variant, the \emph{stochastic} DMFT, which is able to describe Anderson localization as well \cite{Dobrosavljevic97,Dobrosavljevic03}. 
\begin{figure}
\begin{center}
\includegraphics[width=0.55\linewidth]{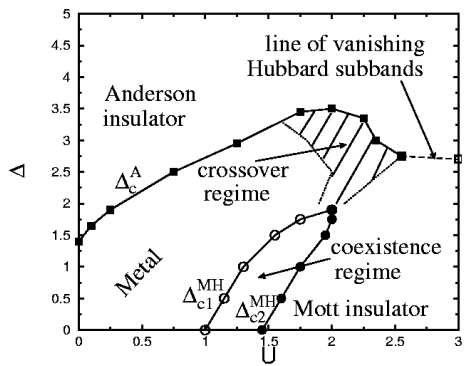}
\end{center}
\caption{\label{fig:8} 
DMFT ground state phase diagram of the disordered Hubbard model in the nonmagnetic phase 
at half filling.   
Figure taken from \cite{anderson-mott}. }
\end{figure}

Within DMFT, the correlated lattice model is mapped onto a self-consistent Anderson impurity 
Hamiltonian 
\bea
H_{\mathrm{SIAM}} &=&\sum_{\sigma }(\epsilon -\mu )c_{\sigma }^{\dagger
}c_{\sigma }+Un_{\uparrow }n_{\downarrow }  \label{2} \\
&&+\sum_{\mathbf{k}\sigma } \left( V_{\mathbf{k}}c_{\sigma }^{\dagger }a_{\mathbf{k}%
\sigma }+V_{\mathbf{k}}^{\ast }a_{\mathbf{k}\sigma }^{\dagger }c_{\sigma } \right)
+\sum_{\mathbf{k}\sigma }\epsilon _{\mathbf{k}}a_{\mathbf{k}\sigma
}^{\dagger }a_{\mathbf{k}\sigma }  \nonumber
\eea
where a single correlated lattice site now constitutes the impurity with a random onsite energy $\epsilon$, 
and the fermions $a_{k \sigma}$ represent a fictitious conduction band with parameters $V_k$ and 
$\epsilon_k$ which have to be determined self-consistently. The chemical potential $\mu  = -U/2$ ensures half filling.  
This effective single-impurity model is solved using Wilson's numerical renormalization group 
\cite{wilson,Costi94,Bulla98,Hofstetter00}. 
Within the stochastic DMFT \cite{Dobrosavljevic03} the self-consistency loop involves 
a geometric disorder average of the local density of states 
$ \rho _{\mathrm{geom}}(\omega )=\exp \left[ \langle \ln \rho _{i}(\omega )\rangle \right] $
which then determines the hybridization function 
$\eta (\omega )=\sum_{\mathbf{k}}|V_{\mathbf{k}}|^{2}/\left( \omega -\epsilon_{\mathbf{k}}\right) $
for the next iteration. For more details see \cite{anderson-mott}. 

The resulting zero temperature phase diagram as a function of 
disorder $\Delta$ and interaction $U$ is shown in Fig.~\ref{fig:8}. 
For weak interaction and disorder the atoms are in a Fermi liquid state (``metal''). 
There are two different metal-insulator transitions: a Mott-Hubbard transition 
for increasing interation $U$, and an Anderson localization transition 
as a function of $\Delta$. Our results indicate that 
the two insulating phases are adiabaticly connected. 
Note, however, that in our DMFT calculation we have so far considered only the paramagnetic phase. 
For non-frustrated lattices (e.g. simple cubic) it is known that an antiferromagnetic instability 
occurs in the pure Mott state. We are currently analyzing how far this antiferromagnetic phase  
extends into the disordered Mott-Anderson insulator \cite{byczuk05}. 
Let us briefly comment on the detection of these different phases. Itinerant versus insulating  
behavior can be identified by a time-flight measurement as in \cite{esslinger}. 
In the Fermi liquid state, delocalization of fermions across the lattice leads to an interference pattern 
which vanishes once the atoms become localized. In order to distinguish the antiferromagnetic Mott insulator 
from the paramagnetic Anderson insulator one could apply spin-resolved Bragg scattering. 

Optical lattices are a promising tool to simulate the above phase diagram experimentally 
since, in contrast to solids, both parameters $U$ and $\Delta$ can be tuned arbitrarily. 
In particular, measurements can be done both in two and three spatial dimensions,  
possibly detecting qualitatively new physics in $d=2$ where DMFT is no longer expected 
to be a good approximation.

\section{Summary and Outlook \label{sec:5}}
In this review we have presented some theoretical aspects of strongly correlated atoms  
in optical lattices. We have shown that these systems can be used to create analogues of 
well established solid-state quantum phases, but with much higher tunability of the model parameters. 
More generally, ultracold atoms can be used to perform quantum simulations of solid-state Hamiltonians 
like the 2d Hubbard model which may be relevant for  high-temperature superconductivity. 
As another example for such a simulation we have discussed interacting fermions with disorder. 
Within a DMFT calculation we observe remarkable re-entrance into the 
itinerant phase due to competing Mott- and Anderson-localization. 
Finally, we have demonstrated that it is possible to use the highly degenerate internal states 
of cold atoms to create new exotic quantum states which have no analogue in condensed matter 
physics. Bosons with multiple spin states can be used to create tunable spin hamiltonians. 
Most prominently, we have discussed a new fermionic SU(3) triplet superfluid state which 
is a toy model for QCD at weak to intermediate interactions. 
Experimental realization of these quantum phases is within reach and 
could significantly increase our understanding of the many-body model systems involved. 

\section{Acknowledgments}

The author would like to thank 
E. Altman, B. Byczuk, I. Cirac, E. Demler, C. Honerkamp, M.D. Lukin, D.~Vollhardt, and P. Zoller 
for collaborations, and I. Bloch, M. Greiner, M. Zwierlein and W. Ketterle for discussions.



\end{document}